\documentclass[aps, prl,  tightenlines, notitlepage, nofootinbib, twocolumn, letterpaper]{revtex4-1}
\pdfoutput=1
\usepackage{epsfig,amsfonts,mathrsfs,amsmath,amssymb,graphicx,color,slashed,bbm}
\usepackage{amsmath,latexsym,amssymb,graphicx,slashed,hyperref,color,enumerate,url,etoolbox}
\hypersetup{colorlinks,citecolor= niceblue,linkcolor= niceblue, urlcolor=niceblue}
\definecolor{niceblue}{rgb}{0.1,0.2,0.6}
\usepackage{comment}

\begin{document}

\title{\Large Neutrino As The Dark Force}
\author{Nicholas Orlofsky}
\author{Yue Zhang}
\affiliation{Department of Physics, Carleton University, Ottawa, Ontario K1S 5B6, Canada}

\begin{abstract}
We point out a novel role for the Standard Model neutrino in dark matter phenomenology where the exchange of neutrinos generates a long-range potential between dark matter particles. The resulting dark matter self interaction could be sufficiently strong to impact small-scale structure formation, without the need of any dark force carrier. This is a generic feature of theories where dark matter couples to the visible sector through the neutrino portal. It is highly testable with improved decay rate measurements at future $Z$, Higgs, and $\tau$ factories, as well as precision cosmology.
\end{abstract}

\maketitle

Dark matter is a key ingredient throughout the evolution of our universe, yet its identity remains unknown. 
Currently, the nature of dark matter is under careful scrutiny at various experimental frontiers
from laboratories to the cosmos, and some hints already exist. Non-gravitational self interaction of 
dark matter could compete with gravity and impact the formation of structures.
Such a new force can help alleviating tensions between numerical simulations and the observed small-scale structure of the universe, known as
the ``core-cusp'' and ``too big to fail'' problems~\cite{Spergel:1999mh, Tulin:2017ara, Bullock:2017xww}.
It could also yield important consequences such as seeding supermassive black hole formation~\cite{Balberg:2002ue,Choquette:2018lvq,Essig:2018pzq,Feng:2020kxv,Xiao:2021ftk}.
The dynamics of self-interacting particle dark matter have been explored in a broad range of theories~\cite{Buckley:2009in,Aarssen:2012fx,Tulin:2013teo,Bellazzini:2013foa,Boddy:2014yra,Hochberg:2014kqa,Soni:2016gzf,Zhang:2016dck,Blennow:2016gde,McDermott:2017vyk,Chu:2018fzy,Chu:2018faw,Chu:2019awd,Agrawal:2020lea,Tsai:2020vpi}, which typically 
host more degrees of freedom than the dark matter itself.
Light dark force carriers are often introduced to mediate the dark matter self interaction
whose potential imprint on the visible universe is tightly constrained~\cite{Bjorken:2009mm,Kaplinghat:2013yxa,Yang:2021adi,Bringmann:2016din,Zhang:2015era}.
A separate small-scale challenge, known as the ``missing satellite'' problem~\cite{Moore:1999nt,2010arXiv1009.4505B,Weinberg:2013aya,Gilman:2019nap}, favors warm dark matter candidates that can erase heretofore-unobserved small structures~\cite{Abazajian:2001nj,Asaka:2005an,Asaka:2006ek,Boyarsky:2009ix,Nemevsek:2012cd,Dror:2020jzy,Bertoni:2014mva}.
Although these puzzles might be relaxed with known physics such as baryonic feedback~\cite{Read:2004xc,Mashchenko:2007jp,Governato:2012fa,Sawala:2015cdf,Wetzel:2016wro,Fattahi:2016nld}, 
they serve as good motivations for building and testing novel dark matter models.

Given that neutrinos are the lightest known particles other than the photon and their properties remain to be fully understood, it is natural to speculate on the possible role of neutrinos to address the above puzzles.
In this article, we demonstrate that dark matter self interactions can be mediated exclusively by Standard Model (SM) neutrinos, without the introduction of dark force carriers. This is a generic possibility within the class of neutrino portal theories. There are several attractive outcomes. First, dark matter self interaction
proceeds via the exchange of two neutrinos. At separations shorter than the inverse neutrino mass,
the potential governing dark matter self interaction is long range, of the form $1/r^5$. 
For asymmetric dark matter, the interaction is repulsive, and the low-energy scattering can be solved within quantum mechanics, independent of short distance physics.
Second, the dark matter-neutrino interaction establishes a thermal history of the dark states and allows robust constraints to be set on their mass scale.
It could also keep the two species in kinetic equilibrium for an extended period, enabling the warm dark matter scenario.
Last but not least, unlike many dark sector models that are secluded from the visible sector, 
the dark matter candidate considered here must interact with known particles through neutrinos.
It is highly testable by precision measurements of the $Z$ boson, Higgs boson, and $\tau$ lepton decay rates. Laboratory and cosmological measurements together provide complementary future probes of such a novel target.

Our starting point is the effective interacting Lagrangian
\begin{equation}\label{1}
\mathcal{L}_\text{int} = \frac{(\bar L_\alpha H) (\chi \phi)}{\Lambda_\alpha} + {\rm h.c.} \ ,
\end{equation}
where $L_\alpha\, (\alpha=e, \mu, \tau)$ is a SM lepton doublet in the flavor basis and $H$ is the Higgs doublet. The dark fermion $\chi$ and scalar $\phi$ are SM gauge singlets but charged under a global $U(1)$ or $Z_2$ symmetry. The lighter is stable and serves as the dark matter candidate, which we assume to be $\chi$ hereafter.
The operator is dimension five, having a cutoff scale $\Lambda_\alpha$.
Interestingly, this neutrino portal operator has been introduced and explored for a number of other motivations~\cite{Falkowski:2009yz,Bertoni:2014mva,Ko:2014bka,Batell:2017rol,Berryman:2017twh,Batell:2017cmf,Becker:2018rve,Folgado:2018qlv,Lamprea:2019qet,Zhang:2020nis}.
Below the electroweak symmetry scale, a Yukawa interaction is generated between the neutrino and dark particles,
\begin{equation}\label{2}
\mathcal{L}_\text{int} = y_\alpha \bar \nu \chi \phi + {\rm h.c.}\ ,
\end{equation}
where $y_\alpha = v/\sqrt{2}\Lambda_\alpha$, and $v=246\,$GeV is the vacuum expectation value of the Higgs field.

\begin{figure}[t]
	\centerline{\includegraphics[width=0.15\textwidth]{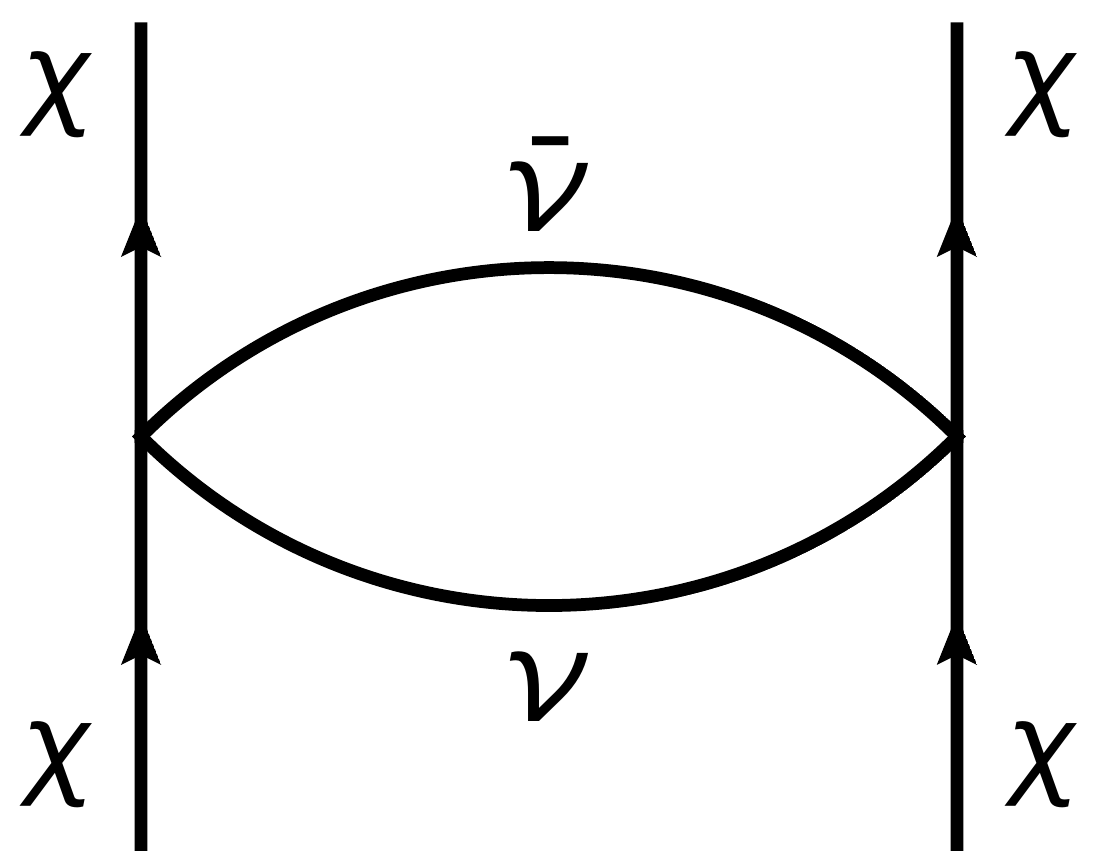}}
	\caption{Two-neutrino exchange diagram that can generate a long-range potential between dark matter particles.}\label{fig:DM-SI}
\end{figure}

{\it Dark matter self-interaction.} \ We first explore dark matter self interaction of relevance to structure formation. Because the dark matter particles are already non-relativistic when they self interact in galaxies and clusters, 
the heavier $\phi$ field could be integrated out from Eq.~\eqref{2}, leading to
\begin{equation}\label{3}
\mathcal{L}_\text{int} \simeq \frac{|y_\alpha|^2}{2(m_\phi^2 - m_\chi^2)} (\bar \chi \gamma^\mu \mathbb{P}_R \chi) (\bar \nu \gamma_\mu \mathbb{P}_L \nu) \ ,
\end{equation}
where $\mathbb{P}_{L,R}=(1\mp\gamma_5)/2$ and $m_{\chi,\phi}$ are the masses of the dark states. 
This effective Lagrangian is obtained by assuming the mass difference between $\phi$ and $\chi$ is much larger than the energy/momentum transfer carried by neutrinos.
The mass square difference factor downstairs captures an enhancement effect when the dark state masses are near.
We first evaluate the $\chi\chi\to \chi\chi$ scattering which occurs via two neutrino exchange at one-loop level (Fig.~\ref{fig:DM-SI}). After non-relativistic reduction, the matrix element takes the form
\begin{equation}\label{4}
i \mathcal{M}(t) = \frac{- i |y_\alpha|^4}{24\pi^2 (m_\phi^2 - m_\chi^2)^2} \left( \frac{1}{\epsilon} + \log\frac{\mu^2}{- t} + \frac{5}{3} \right) t \ ,
\end{equation}
where $t=q^2<0$ with $q^\mu$ the four-momentum transfer and $\mu$ is the renormalization scale.
The UV divergence ($\epsilon \to 0$) is regularized in the full theory (including $\phi$).

The long-range part of the potential after resumming the multiple two-neutrino exchange can be derived using the dispersion theory technique~\cite{Feinberg:1968zz,Feinberg:1989ps,Hsu:1992tg},
\begin{equation}\label{5}
\begin{split}
V(r) &= \int \frac{d^3 q}{(2\pi)^3} \frac{-1}{2\pi i} \int_0^\infty dt' \frac{{\rm disc} \mathcal{M}(t') }{t'-q^2} \\
&= \frac{|y_\alpha|^4}{128\pi^3 (m_\phi^2 - m_\chi^2)^2 r^5} \ ,
\end{split}
\end{equation}
where ${\rm disc} \mathcal{M}(t')$ is the discontinuity of $\mathcal{M}$ across its branch cut in the complex plane of $t'$.
A similar potential from neutrino exchange within the SM has also been explored \cite{Stadnik:2017yge,Thien:2019ayp,Ghosh:2019dmi,Segarra:2020rah,Bolton:2020xsm}.
At very short distances where $\phi$ cannot be integrated out, and at very large distances where neutrino masses are non-negligible,
the $1/r^5$ form of the above potential will break down~\cite{Grifols:1996fk}.
Thanks to the repulsive nature of the potential and the non-relativistic nature of dark matter scattering considered here, the probability for two dark matter particles to find each other at distances $r\lesssim m_\phi^{-1}$ is highly suppressed. 
This motivates us to proceed by assuming only $\chi$ is present in the universe, which coincides with the asymmetric dark matter idea~\cite{Nussinov:1985xr}.~\footnote{Had we considered symmetric dark matter, the $\chi\bar\chi$ interaction potential would be attractive and require detailed knowledge of short distance physics~\cite{Lepage:1997cs}.}
In addition, the potential energy at $r\gtrsim m_\nu^{-1}$ (inverse of neutrino mass) is much lower than the typical dark matter kinetic energy in galaxies and clusters.
Thus, it is a good approximation to simply consider the potential in Eq.~(\ref{5}) for the discussions here.

\begin{figure*}
\centerline{
\includegraphics[width=1\columnwidth]{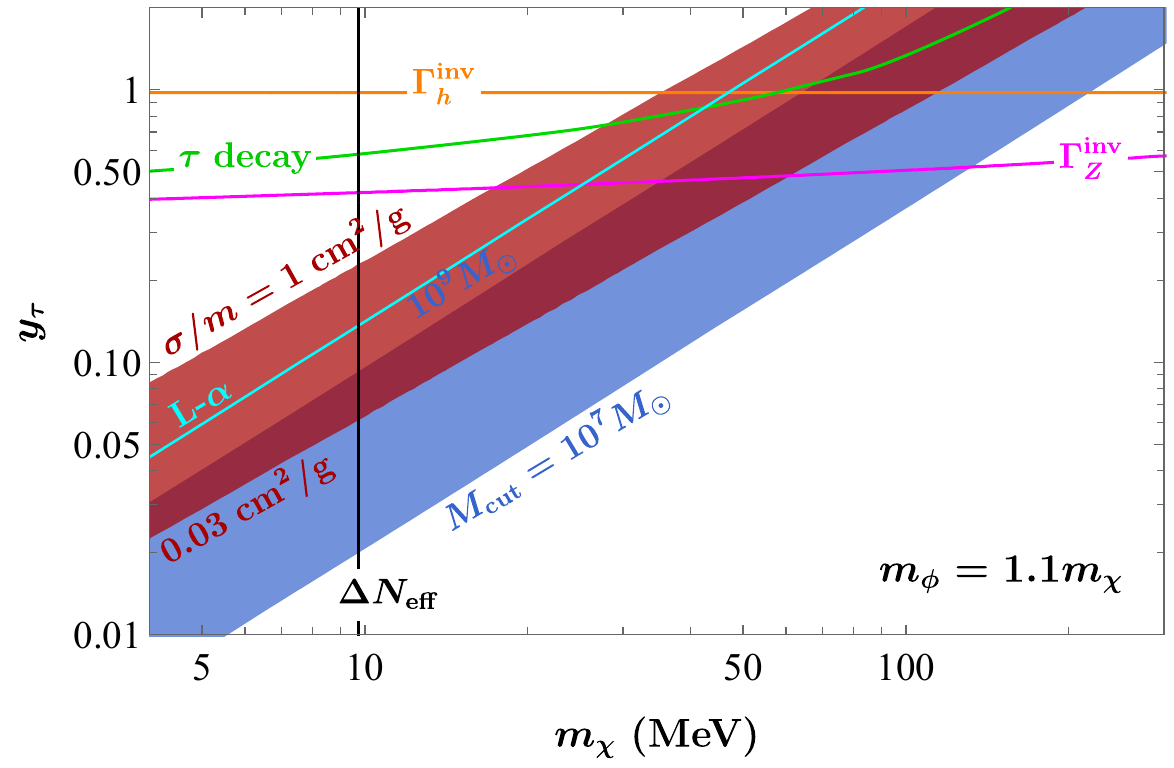}
\includegraphics[width=1\columnwidth]{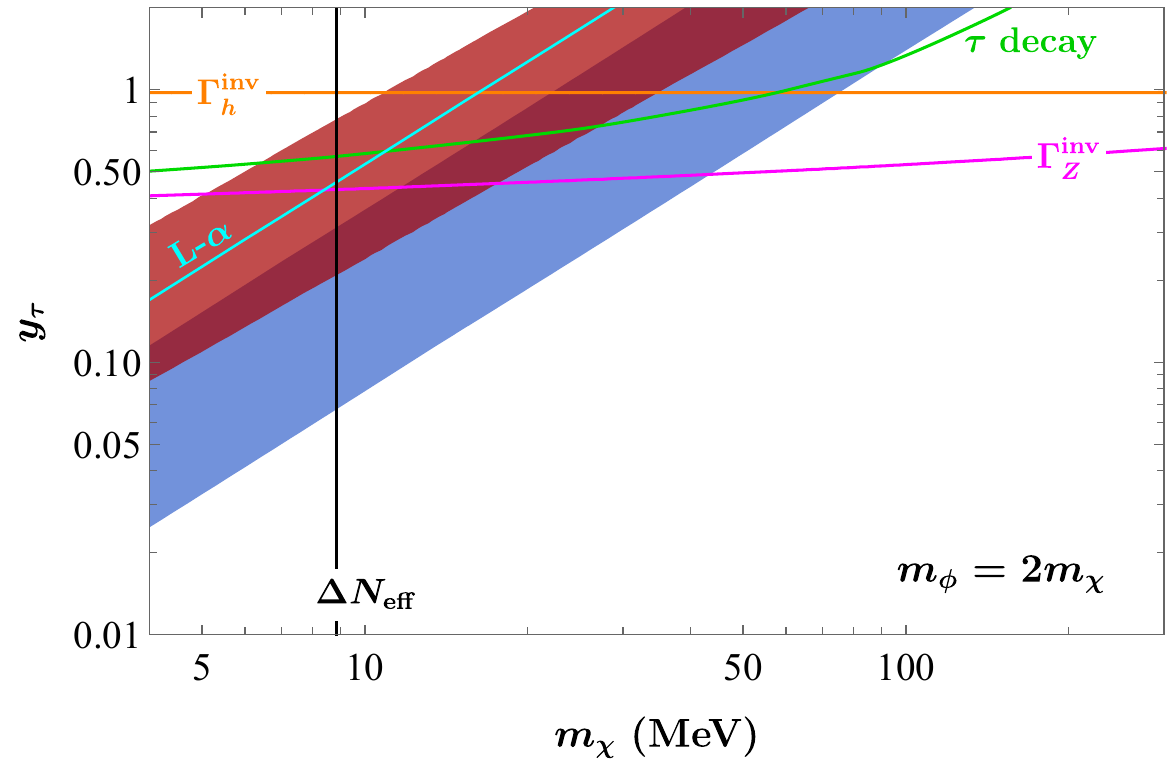}}
\caption{The region of parameter space where the long-range force due to neutrino exchange can generate a sufficiently large dark matter self-interaction cross section for addressing puzzles in small-scale structure formation. The dark red shaded band corresponds to $0.03\,{\rm cm^2/gram}\leq \sigma_{\chi\chi\to\chi\chi}/m_\chi \leq 1\,{\rm cm^2/gram}$. 
The dark blue band corresponds to $\chi$ serving as a warm dark matter. The two regions intersect in the darkest shaded region.
The cyan curve sets an upper bound on neutrino-$\chi$ interactions from Lyman-$\alpha$.
The lower bound on the dark matter mass set by $\Delta N_{\rm eff}$ is shown by the vertical black line.
Existing upper bounds on the Yukawa coupling $|y_\tau|$ are set from the invisible decays of the $Z$ boson (magenta curve) and Higgs boson (orange curve) and the leptonic decay of the $\tau$ (green curve).
All bounds are at 95\% confidence level.
}\label{fig1}
\end{figure*}

The low energy scattering problem for a repulsive $1/r^5$ potential is well defined in quantum mechanics. It is free from ultraviolet (UV) dependence in spite of being singular~\cite{Pais:1964zz,RevModPhys.43.36}. In Refs.~\cite{1965NCimS..38..443D, 1965NCimA..40..739G}, the analytic expression of the scattering phase shift has been derived for all partial waves.
In particular, the $S$-wave phase shift takes the form
\begin{equation}\label{6}
\begin{split}
\tan \delta_0 &= \frac{3^{-2/3} \Gamma(-1/3)}{\Gamma(1/3)} f^{5/3} + \mathcal{O}(f^5) \ , \\
f &= k^{3/5} \left[\frac{\mu_{\chi\chi} |y_\alpha|^4}{64\pi^3 (m_\phi^2 - m_\chi^2)^2}\right]^{1/5} \ ,
\end{split}
\end{equation}
where $\Gamma$ is the Euler gamma function, $\mu_{\chi\chi}=m_\chi/2$ is the reduced mass of the $\chi\chi$ system, and $k$ is the relative momentum of scattering.
For non-relativistic dark matter and perturbative values of $y_\alpha$, we find the expansion parameter $f\ll1$. Thus, the $f^{5/3}$ term dominates.
Higher partial wave ($\ell \geq1$) phase shifts begin from order $f^5$ or higher and are not important. 
The scattering cross section is $S$-wave dominated and well approximated by
\begin{equation}\label{7}
\sigma_{\chi\chi\to\chi\chi} \simeq \frac{4\pi}{k^2} \sin^2\delta_0 \simeq 0.027 \left[\frac{m_\chi |y_\alpha|^4}{(m_\phi^2 - m_\chi^2)^2}\right]^{2/3} \ .
\end{equation}
The resulting cross section is insensitive to the relative velocity as long as $f\ll1$.
This implies the same prediction of the dark matter self-interaction cross section applies to various astrophysical objects, from dwarf galaxies to clusters.
It is important to note that the Born approximation does not work. Indeed, the resulting cross section goes as $|y_\alpha|^{8/3}$ rather than $|y_\alpha|^8$, indicating the importance of resumming multiple neutrino bubble exchange contributions~\cite{1965NCimS..38..443D}.
This is the key for generating a sizable dark matter interaction in spite of the loop-level origin of the potential, Eq.~\eqref{5}.
Numerically, we have verified the above results using the shooting method.

With the cross section in Eq.~\eqref{7}, we derive the parameter space for self-interacting dark matter. In Fig.~\ref{fig1}, the red shaded band corresponds to 
$0.03\,{\rm cm^2/gram}\leq \sigma_{\chi\chi\to\chi\chi}/m_\chi \leq 1\,{\rm cm^2/gram}$, potentially relevant for solving the various small-scale puzzles~\cite{Kaplinghat:2015aga}.
The upper bound is set by the Bullet Cluster observation~\cite{Randall:2007ph,2015Sci...347.1462H,Robertson:2016xjh}.
Here we focus on the coupling of the $\nu_\tau$ neutrino with the dark sector, which receives the least constraints compared to other flavor choices (see discussions below).
In the left and right panels, we choose the mass ratios $m_\phi/m_\chi=1.1$ and $2$, respectively.
When $m_\phi$ and $m_\chi$ are closer, the effective coupling in Eq.~\eqref{3} is more enhanced, allowing for smaller values of $y_\alpha$.

{\it Overlap with warm dark matter.} \ The interaction of Eq.~\eqref{2} has another significant cosmological implication. 
A sufficiently large $y_\alpha$ can keep dark matter and neutrinos in kinetic equilibrium with each other for an extended period of time, leading to a
suppressed dark matter density power spectrum at small length scales via collisional damping. The cutoff mass scale of the smallest gravitationally bound dark matter halo is~\cite{Bertoni:2014mva} 
\begin{equation}\label{7.1}
M_{\rm cut} \simeq 10^8 M_\odot \left[ \frac{|y_\alpha|}{0.3} \right]^3 \left[\frac{20\,\rm MeV}{m_\chi} \right]^{\frac{3}{4}} \left[ \frac{26\,\rm MeV}{\sqrt{m_\phi^2 - m_\chi^2}} \right]^{3} \ ,
\end{equation}
where $M_\odot$ is the solar mass.
With $10^7M_\odot \lesssim M_{\rm cut} \lesssim 10^{9}M_\odot$, $\chi$ is a warm dark matter candidate and can shed light on the ``missing satellite'' problem. The favored parameter space is depicted by the 
blue shaded band in Fig.~\ref{fig1}.
Remarkably, there is an overlap with the self-interacting dark matter region derived above, allowing all puzzles in small-scale structure formation to be tied to this simple framework.

The neutrino-dark matter interaction can also be constrained by larger scale probes, including the cosmic microwave background (CMB) \cite{Escudero:2015yka,DiValentino:2017oaw,Diacoumis:2018ezi} and large scale structure \cite{Wilkinson:2014ksa,Campo:2017nwh}. The constraint from Lyman-$\alpha$ \cite{Wilkinson:2014ksa} is the strongest among these, setting an upper limit on the elastic scattering cross section of $\sigma_\text{el}/m_\chi < 10^{-36}\, \text{cm}^2/\text{MeV}$ for neutrino energies of around 100 eV. This bound is shown by the cyan curve in Fig.~\ref{fig1}. Other constraints from higher energy neutrinos, {\it e.g.}, those detected from SN1987A \cite{Mangano:2006mp} or at IceCube \cite{Arguelles:2017atb,Kelly:2018tyg,Yin:2018yjn,Choi:2019ixb}, do not set a bound on the plotted parameter space.

{\it Early universe constraints.} \ To derive the above self interaction results, we have made the assumption that dark matter is asymmetric. This has the advantage of making the dark matter self interaction repulsive, thus the low energy observables are free from UV dependence. Here we show the strength of the dark matter-neutrino interaction is compatible with such an assumption.
When the temperature of the universe is higher than the mass of $\chi$ and $\phi$, the Yukawa interaction of Eq.~\eqref{2} necessarily thermalizes them with neutrinos.
The key cross section for annihilating away the $\bar\chi$ particles is
\begin{equation}\label{7.5}
(\sigma v_{\rm Møl})_{\chi\bar\chi\to\nu\bar\nu} = \frac{|y_\alpha|^4 m_\chi^2}{32\pi (m_\chi^2 + m_\phi^2)^2} \ .
\end{equation}
The annihilation of $\phi \phi^* \to\nu\bar\nu$ is $P$-wave suppressed.
For $m_\phi \sim m_\chi$, the condition for efficiently depleting the symmetric population (i.e., $(\sigma v_{\rm Møl})_{\chi\bar\chi\to\nu\bar\nu} \gg 3 \times 10^{-26}\,\text{cm}^3/\text{s}$) corresponds to
\begin{equation}\label{7.6}
|y_\alpha| \gg 0.004 \left( \frac{m_\chi}{10\,\rm MeV} \right)^{1/2}  \ .
\end{equation}
Clearly, this requirement is easily satisfied for the coupling values of interest in Fig.~\ref{fig1}.

There is an important constraint on the lightness of dark matter from $\Delta N_{\rm eff}$, the excess radiation degrees of freedom in the universe, during the Big Bang nucleosynthesis and recombination epochs~\cite{Nollett:2014lwa,Escudero:2018mvt,Pitrou:2018cgg}.
To support a $U(1)$ stabilizing symmetry for the  dark sector, we must assume $\chi$ is a Dirac fermion and $\phi$ a complex scalar. 
For the two mass ratios considered in Fig.~\ref{fig1}, lower limits on the $\chi$ mass are 9.7 and 8.9\,MeV, respectively. Here, we apply a conservative $2\sigma$ limit $\Delta N_\text{eff} < 0.5$ \cite{Aghanim:2018eyx,Riess_2018}.  
The upcoming CMB Stage-IV experiment~\cite{Abazajian:2016yjj} 
can probe the dark matter mass up to $16.9$\,MeV for $m_\phi \gg m_\chi$ \cite{Escudero:2018mvt}.

{\it Laboratory constraints.} \ The interaction strength of the neutrino portal to the dark sector can be probed by a number of precision measurements of known particles.
The effective operator of Eq.~\eqref{1} contributes to the invisible decay width of the Higgs boson,
\begin{equation}\label{8}
\begin{split}
&\Gamma_{h\to \bar\nu_\alpha \chi \phi} = \frac{|y_\alpha|^2 G_F m_h^3}{256\sqrt{2} \pi^3}
\int_{(\sqrt{z_\chi}+\sqrt{z_\phi})^2}^1 dx f_h(x, z_\chi, z_\phi)\ , \\
&f_h(x, z_\chi, z_\phi) = x^{-2} (1-x)^2 (x-z_\phi + z_\chi) \\
&\hspace{2cm} \times \sqrt{x^2 -2 x(z_\chi +z_\phi)+(z_\chi -z_\phi)^2} \ ,
\end{split}
\end{equation}
where $z_\chi = m_\chi^2/m_h^2$ and $z_\phi = m_\phi^2/m_h^2$.
The $f_h$ integral evaluates to $1/3$ in the limit $z_\chi=z_\phi=0$.
An upper bound on $y_\alpha$ is derived by requiring the branching ratio of this decay (adding the charge-conjugation channel) to be less than 24\% at 95\% confidence level~\cite{Khachatryan:2016whc,Aaboud:2019rtt}, as shown by the
horizontal orange curve in Fig.~\ref{fig1}. An optimistic projected sensitivity for the Higgs invisible decay branching ratio at the HL-LHC of 3\%~\cite{CMS:2017cwx} would strengthen this limit to $|y_\alpha|<0.3$.

The Yukawa interaction of Eq.~\eqref{2} contributes to the invisible decay width of the $Z$ boson, 
\begin{equation}\label{9}
\begin{split}
&\Gamma_{Z\to \bar\nu_\alpha \chi \phi} = \frac{|y_\alpha|^2 G_F m_Z^3}{768\sqrt{2} \pi^3}
\int_{(\sqrt{z_\chi}+\sqrt{z_\phi})^2}^1 dx f_Z(x, z_\chi, z_\phi)\ , \\
&f_Z(x, z_\chi, z_\phi) = x^{-1} (2+x) f_h(x, z_\chi, z_\phi) \ ,
\end{split}
\end{equation}
where here $z_\chi = m_\chi^2/m_Z^2$ and $z_\phi = m_\phi^2/m_Z^2$.
The existing constraint on the $Z$ boson invisible width~\cite{Zyla:2020zbs} sets an upper bound on $y_\alpha$ as shown by the magenta curve in Fig.~\ref{fig1}.
Both the Higgs and $Z$ decay constraints apply universally to all neutrino flavors $\alpha=e, \mu, \tau$.

For $\alpha=\tau$, the Yukawa interaction of Eq.~\eqref{2} leads to a new decay mode $\tau^- \to \mu^- \bar \nu_\mu \chi \phi$ which mimics the normal leptonic decay.
A similar process was considered in~\cite{Brdar:2020nbj}. We simulate this four-body decay with FeynRules~\cite{Alloul:2013bka} and MadGraph~\cite{Alwall:2014hca} and obtain an upper bound on $y_\tau$ as a function of the dark matter mass,
as shown by the green curve in Fig.~\ref{fig1}.
For other flavor choices $\alpha=\mu, e$, much stronger constraints arise from leptonic decays of charged kaons and pions.
In those cases, the parameter space of interest to cosmology has already been excluded.

{\it Neutrino self-interaction.} \ The neutrino portal coupling could also lead to non-standard neutrino self interaction, which arises from a box diagram with $\chi$ and $\phi$ in the loop.
The lower energy effective operator takes the form
\begin{equation}
\begin{split}
\mathcal{L}_{\rm SI\nu} &= G_{\rm eff} (\bar \nu \gamma^\mu P_L \nu) (\bar \nu \gamma_\mu P_L \nu) \ ,  \\
G_{\rm eff} &= \frac{|y_\alpha|^4 \left(m_\phi^4 - m_\chi^4 - 2 m_\phi^2 m_\chi^2 \log \frac{m_\phi^2}{m_\chi^2} \right)}{64\pi^2 (m_\phi^2 - m_\chi^2)^3} \ .
\end{split}
\end{equation}
In the limit $m_\phi \simeq m_\chi$, $G_{\rm eff} \simeq |y_\alpha|^4/(192 \pi^2 m_\chi^2)$. 
A sizable neutrino self interaction has been suggested as an ingredient for solving the Hubble tension~\cite{Kreisch:2019yzn,Oldengott:2017fhy}.
However, the relevant parameter regions are already ruled out by laboratory and $\Delta N_\text{eff}$ constraints in this model. 

{\it UV completion.}\ For completeness, we briefly discuss a UV complete model for the effective operator in Eq.~\eqref{1}.
It can be generated by integrating out a gauge singlet vectorlike fermion that couples to both the visible and dark sectors. The interacting Lagrangian takes the form
$\mathcal{L}_{\rm UV} = \lambda_{V} \bar L_\alpha H N_R + \lambda_D \bar N_L \chi \phi + M \bar N_L N_R +{\rm h.c}$. The first Yukawa term allows for a heavy-light neutrino mixing below the electroweak scale,
$N_L = \sqrt{1-|U_{\alpha4}|^2} \hat N_L + U_{\alpha4} \hat\nu_\alpha$, where the hat fields are physical states and $U_{\alpha 4} = \lambda_V v/\sqrt{M^2+\lambda_V^2 v^2}$. Together with the second Yukawa term, 
we obtain the relationship, $y_\alpha = \lambda_D U_{\alpha4}$, where $y_\alpha$ is the Yukawa coupling introduced in Eq.~\eqref{2}.
In this UV completion, there are additional constraints on the mixing parameter $U_{\alpha4}$.
For $\alpha=\tau$ flavor and $N$ mass around/above the electroweak scale, the strongest constraints are from $\tau$ lepton decays, $|U_{\tau4}|\lesssim 0.2$~\cite{deGouvea:2015euy, Batell:2017cmf}.
If $|U_{\tau4}|$ is close to this upper bound, the favored range of $y_\tau$ in Fig.~\ref{fig1} can be obtained with an order one fundamental Yukawa coupling $\lambda_D$.

{\it Other possibilities.}\ If the mass difference between $\phi$ and $\chi$ 
is tuned to be comparable to the typical kinetic energy of dark matter, the effective Lagrangian Eq.~\eqref{3} will break down. 
In galaxies, $\phi$ particles could be produced on shell via $\chi\chi$ collisions.
In this case, $\phi$ could still decay quickly back to $\chi$ and a neutrino within the cosmological time scale, leading to dissipative dark matter of the inelastic kind~\cite{Das:2017fyl,Vogelsberger:2018bok}.
In the limiting case where $\phi$ and $\chi$ are degenerate, both will serve as dark matter. The Born level momentum-transfer cross section~\cite{Tulin:2013teo} of $\chi-\phi$ scattering via neutrino exchange is
$\sigma_{\chi\phi\to \chi\phi} \simeq |y_\alpha|^4/(32\pi m_\chi^2 v^2)$ in the limit $m_\chi v \gg m_\nu$, where $v$ is the relative velocity.
The Bullet Cluster constraint on this cross section implies $|y_\alpha| < 0.026 (m_\chi/10\,{\rm MeV})^{3/4} (v/10^{-3})^{1/2}$. 
This constraint could be affected by non-perturbative effects due to multiple neutrino exchanges. 

So far, we have assumed dark matter to be asymmetric so that the neutrino mediated potential, Eq.~(\ref{5}), is repulsive.
For symmetric dark matter, the potential would be attractive and could impact dark matter annihilation and indirect detection.
Similar physics as discussed above can be generalized to dark matter self interaction via the exchange of other motivated light fermions, such as the sterile neutrino~\cite{Foot:1995pa,Berezhiani:1995yi,Chun:1995bb,Dvali:1998qy,Ghosh:2010hy,Kopp:2013vaa}.

{\it Summary.} \ The findings of this work demonstrate that the simple neutrino portal offers a rich dark matter phenomenology.
Self-interacting dark matter can occur without introducing dark force carriers, but rather via the exchange of Standard Model neutrinos.
It generates a $1/r^5$ potential between dark matter at distances shorter than the inverse of the neutrino mass.
We identify a parameter space where dark matter has sufficiently strong self interactions to influence small-scale structure formation.
Meanwhile, the interaction between dark matter and neutrinos could accommodate a warm dark matter candidate.
This interplay allows for a unified solution to all the puzzles in small-scale structure formation.
The corresponding neutrino portal interaction is a well-motivated target  
for precision measurements of decay rates of the $Z$ boson, Higgs boson, and $\tau$ lepton at a number of future collider experiments~\cite{Baer:2013cma,Bondar:2013cja,Kou:2018nap,Blondel:2018mad,CEPCStudyGroup:2018ghi,Cepeda:2019klc}, 
as well as precision measurement of  $\Delta N_{\rm eff}$ by the upcoming CMB-S4 project.

We thank James Cline, Walter Tangarife, Hai-Bo Yu, and Yi-Ming Zhong for useful discussions and communications. 
This work is supported by the Arthur B. McDonald Canadian Astroparticle Physics Research Institute.

\bibliography{References}

\end{document}